\documentstyle[psfig,twocolumn,prc,aps]{revtex}
\begin{document}
\draft

\title{Gamow Teller strength in $^{54}$Fe and $^{56}$Fe}

\author{E.  Caurier$^{1}$\thanks{E-mail: caurier@crnhp4.in2p3.fr}, A.
  Poves$^{2}$\thanks{E-mail: poves@nucphys1.ft.uam.es}, A.  P.
  Zuker$^{1}$\thanks{E-mail: zuker@crnhp4.in2p3.fr} and
  G.~Mart\'{\i}nez-Pinedo$^{2}$\thanks{E-mail:
    gabriel@nucphys1.ft.uam.es}}

\address{$^{1}$Physique Th\'eorique, B\^at 40/1 CRN,
  IN2P3-CNRS/Universit\'e Louis Pasteur BP 28, F-67037 Strasbourg
  Cedex 2, France}

\address{$^{2}$Departamento de F\'{\i}sica Te\'orica, Universidad
    Aut\'onoma de Madrid, E-28049 Madrid, Spain}

\maketitle

\begin{abstract}
  Through a sequence of large scale shell model calculations, total
  Gamow-Teller strengths ($S_+$ and $S_-$) in $^{54}$Fe and $^{56}$Fe
  are obtained.  They reproduce the experimental values once the
  $\sigma\tau$ operator is quenched by the standard factor of $0.77$.
  Comparisons are made with recent Shell Model Monte Carlo
  calculations.  Results are shown to depend critically on the
  interaction.  From an analysis of the GT+ and GT$-$ strength
  functions it is concluded that experimental evidence is consistent
  with the $3(N-Z)$ sum rule.
\end{abstract}


The charge exchange reactions $(p,n)$ and $(n,p)$ make it possible to
observe, in principle, the total Gamow-Teller strength distribution in
nuclei. The experimental information is particularly rich in $^{54}$Fe
and $^{56}$Fe \cite{vett,romq,rapp,ander,elka} and the availability of
both GT+ and GT$-$ makes it possible to study in detail the problem of
renormalization of $\sigma \tau$ operators.  Moreover, these nuclei are
of particular astrophysical interest \cite{astro}, and they have been
the object of numerous theoretical studies.

In this paper we present the results obtained with the largest shell
model diagonalizations presently possible.  First we concentrate on
the study of the total strengths $S_+$ and $S_-$. After a brief review
of existing calculations, we estimate the exact values in a full
$0\hbar\omega$ space, stressing the need of ensuring the correct
monopole behaviour for the interaction.

The second part of the paper deals with the GT+ and GT$-$ stength
functions.  The analysis will confirm that the ``standard'' quenching
factor of 0.77 is associated to suppression of strength in the
$0\hbar\omega$ model space, but that little strength is actually
``missing'' (i.e., unobserved).

\medskip

\noindent
{\bf I. Total Strengths $S_+$ and $S_-$}

\smallskip

The experimental situation is the following:
\begin{itemize}

\item[i)] $^{54}$Fe$(n,p)^{54}$Mn,\\
  $S_+$ = 3.1$\pm$0.6 from \cite{vett}, strength below 10 MeV.\\ $S_+$
  = 3.5$\pm$0.3$\pm$0.4
  from \cite{romq}, strength below 9 MeV.

\item[ii)]  $^{54}$Fe$ (p,n) ^{54}$Co,\\
    $S_-$ = 7.5$\pm$1.2     from \cite{vett}, strength below 15 MeV.\\
    $S_-$ = 7.8$\pm$1.9     from \cite{rapp}, strength below 13.5 MeV.\\
    $S_-$ = 7.5$\pm$0.7     from \cite{ander}, strength below
    $\approx$ 24 MeV,
    (but see discussion in section II).

\item[iii)]  $^{56}$Fe$ (n,p) ^{56}$Mn,\\
    $S_+$ = 2.3$\pm$0.2$\pm$0.4    from \cite{romq}, strength below 7
    MeV.\\
    $S_+$ = 2.9$\pm$0.3    from \cite{elka}, strength below 8.5 MeV.

\item[iv)]  $^{56}$Fe$ (p,n) ^{56}$Co,\\
    $S_-$ = 9.9$\pm$2.4        from \cite{rapp}, strength below 15 MeV.

\end{itemize}

\medskip

The theoretical approaches include: shell model calculations in the
$pf$ shell with different levels of truncation, RPA, quasiparticle RPA
and Shell Model Monte Carlo (SMMC) extrapolations. Let us examine the
results.

\smallskip

${\bf ^{54}Fe }$.  {\it Previous shell model calculations}.
Throughout the paper $f$ stands for 1$f_{7/2}$ and $r$ for any of the
remaining orbits.  Truncation level $t$ means that a maximum of $t$
particles are promoted from the $1f_{7/2}$ orbit to the higher ones,
i.e., that the calculation includes the following configurations:

$f^{n-n_0} r^{n_0}$, $f^{n-n_{0}-1} r^{n_{0}+1}$, $\cdots \cdots$
$f^{n-n_{0}-t} r^{n_{0}+t}$

with $n=A-40$. $n_0$ is different from zero when more than eight
neutrons (or protons) are present and at $t=n-n_0$ we have the full
space calculation.

Given a choice of $t=t_p$ for a parent state having $n_0=n_{0p}$, to
ensure respect of the $S_--S_+= 3(N-Z)$ sum rule, the truncation level
for daughter states having $n_0=n_{0d}$, must be taken to be
\mbox{$t_d=t_p+1+n_{0p}-n_{0d}$}.

In the simplest case we have $t=0$ i.e., 0p-2h configurations with
respect to the $^{56}$Ni closed shell for the $^{54}$Fe ground state,
and 1p-3h configurations for the $^{54}$Mn daughters. The result
---$S_+$ =10.29 Gamow Teller units--- is independent of the
interaction.  The calculation was extended to $t=1$ by Bloom and
Fuller \cite{bloom}, using the interaction of ref. \cite{bint},
obtaining $S_+ =9.12$.  A similar calculation by Aufderheide {\it et
  al.} \cite{auf} yields $S_+ =9.31$ (interaction from
\cite{vanhe,koops}).  Muto \cite{muto}, using the interaction
\cite{muint}, made a $t=$2-like calculation that did not respect the
$3(N-Z)$ sum rule, but the author estimated the influence of this
violation and proposed $S_+$ =7.4. Finally Auerbach {\em et al.}
\cite{auer} have made a $t=$2 calculation using the interaction,
MSOBEP, fitted in \cite{brint} (BR from now on), and obtain $S_+$
=7.05.

{\it QRPA}. The calculation of Engel {\em et al.} \cite{engel} yields
$S_+ =5.03$, to be compared with QRPA or RPA calculations of Auerbach
\cite{auer} leading to $S_+ =6.70$.

{\it Shell Model Monte Carlo}. The calculation of Alhassid {\em et
  al.} \cite{alha}, in the full $pf$ shell, using the BR interaction
extrapolates to $S_+ =4.32\pm0.24$. (here the error bar includes only
the statistical uncertainties but not those associated to the
extrapolation or to possible sistematic errors of the method).

{\it Large t Shell Model calculations}.  All the previous results
point to a reduction of GT+ strength as correlations are introduced
and to a rather large dispersion of the calculated values depending on
the interaction and the approach used. Therefore, to obtain a reliable
$S_+$ value, the method and the interaction must demonstrate their
ability to cope with a large number of other properties of the region
under study.  Calculations in the $pf$ shell \cite{czpm} using the KB3
interaction ---a minimally modified version of the Kuo Brown G-matrix
\cite{kb}--- fulfill this condition since they give an excellent
description of most of the observables in the region up to A=50. The
same interaction was used years ago in perturbation theory to describe
nuclei up to $^{56}$Ni \cite{pz}, with fair success. It should be
mentioned that the monopole modifications in KB3 involve only the
centroids $V_{ff}$ and $V_{fr}$. The $V_{rr}$ values were left
untouched and may need similar changes.

It is not yet possible to perform a full $pf$ shell calculation in
$^{54}$Fe. However, we can come fairly close by following the
evolution of the total strength as the valence space is increased. The
shell model matrices are built and diagonalized and the GT strengths
calculated with the code ANTOINE \cite{ant}.  Full Lanczos iterations
in spaces that reach maximum m-scheme dimension of $1.4 \cdot 10^{7}$
are necessary for the parent states.  Acting on them with the $\sigma
\tau$ operator to calculate the strength, leads to spaces of
$m$-scheme dimension of $4.1 \cdot 10^{7}$.

In addition to KB3, to compare with the results of the SMMC
extrapolations of \cite{alha}, we have used the BR interaction
\cite{brint}. The results are collected in table I and we proceed to
comment on them.

\begin{enumerate}

\item The $t=$5 calculation should approximate the exact ground state
  energy reasonably well, as can be gathered from the small gain of
  270 KeV achieved when increasing the space from $t=$4 to $t=$5.

\item The SMMC result using the BR interaction, \mbox{$-55.5\pm0.5$
    MeV}, is some 1 MeV above the exact energy since our $t=$5 result
  gives an upper bound. Consequently the SMMC error bars in
  \cite{alha} are underestimated.

\item The result of our $t=$2 calculation using BR differs slightly
  from the one in \cite{auer} (7.22 {\it vs.} 7.05). This is due to the
  readjustment of the single particle energies made in \cite{auer}
  with respect to the values of \cite{brint}.

\item Auerbach {\em et al.} proposed an extrapolation of their $t=2$
  calculation to the full space, based on the behaviour of $S_+$ in
  $^{26}$Mg as a function of \mbox{B(E2; $2^{+} \rightarrow 0^{+}$)}.
  Although it is true that there is a qualitative correlation between
  these two observables (the bigger the quadrupole collectivity the
  smaller the $S_+$ value), it is difficult to go further and to
  obtain a quantitative prediction.  In figure 1 we have plotted the
  $S_+$ values {\it vs.} B(E2) for the BR interaction and several
  truncations. It is clear that no simple correlation pattern comes
  out. Notice also that the extrapolation in \cite{auer} gives $S_+
  =6.4$ compared to $S_+=5.5$ in the $t=5$ calculation.

\end{enumerate}

Before we discuss the differences between the results of the BR and
KB3 interactions and between shell model diagonalizations and Monte
Carlo extrapolations we examine the situation in $^{56}$Fe.

\smallskip

${\bf ^{56}Fe }$.  Bloom and Fuller made a $t=0$ calculation
\cite{bloom} that yields $S_+ =10.0$ (interaction from \cite{bint}).
Anantaraman {\it et al.} \cite{anan} (interaction from
\cite{vanhe,koops}) obtain $S_+ =9.25$ for $t=0$ and $S_+ =7.38$ for
$t=1$.  The SMMC result \cite{dean} is shown in table I together with
the numbers coming out of several truncations for both KB3 and BR
interactions.

\smallskip

{\bf The influence of the interaction.} The interactions KB3 and BR
lead to different single particle spectra for $^{57}$Ni. The sequence
of levels obtained in the calculations up to $t=4$ are compared with
the experimental data in table II. It is apparent from the table that
the BR interaction places the $1f_{5/2}$ orbit too low.  As a
consequence the dominant configuration in the $^{56}$Fe ground state
predicted by BR is $(1f_{7/2})^{14} (1f_{5/2})^{2}$ instead of
$(1f_{7/2})^{14} (2p_{3/2})^{2}$ as given by KB3.  This explains the
very large difference in $S_+$ values observed in table I, already at
the $t=$0 level: For a pure $(1f_{7/2})^{14} (2p_{3/2})^{2}$
configuration the total strength is 10.3, while for a pure
$(1f_{7/2})^{14} (1f_{5/2})^{2}$ it amounts to only 5.7. In $^{54}$Fe
the situation is not so dramatic because the leading configuration,
$(1f_{7/2})^{14}$, is the same in both cases.  Still the BR value is
20$\%$ smaller than the KB3 one, due to an excess of $1f_{7/2}$ -
$1f_{5/2}$ mixing in the ground state.  From that we conclude that the
BR interaction underestimates the $S_+$ values for nuclei with N or Z
greater than 28.

Table II also shows the valus of the ``gaps'' defined by:

$\Delta= 2BE(^{56}$Ni$)- BE(^{57}$Ni,$\frac{3}{2}^{-})- BE(^{55}$Ni).

The strong staggering between even and odd values of $t$ makes it
difficult to obtain a reliable extrapolation. The overall trend for
the gap is to decrease as $t$ increases. Nevertheless, it is probable
that the exact value for KB3 will remain somewhat larger than the
experimental one. In this case a slight revision of the monopole terms
would be needed.

\medskip

{\bf Shell Model extrapolations.} In figure 1 we show the evolution of
the total GT strength with the level of truncation in $^{54}$Fe ,
$^{56}$Fe and several cases ($^{48}$Ti, $^{50}$Ti, $^{48}$Cr, and
$^{50}$Cr) for which exact results are available. If we continue the
$^{54}$Fe calculated values with lines paralel to the A=50 ones, we
get the following extrapolated values:

$^{54}$Fe ; $S_+$(KB3)= 6.0 ; $S_+$(BR)= 5.0

\noindent If we assume that the value of the difference between the
$t=$4 to $t=$5 result and the exact one is the same in $^{54}$Fe and
$^{56}$Fe, the corresponding extrapolation is

$^{56}$Fe ; $S_+$(KB3)= 4.5

\noindent these values are fully consistent with the experimental
results if we use the standard 0.77 renormalization of the
Gamow-Teller operator (\cite{brwil,oster,cpz} and section II).
For $^{54}$Fe we have

$S_+$(exp)= 3.1$\pm$0.6 ; 3.5$\pm$0.7 {\it vs.}

$S_+$(KB3)= 3.56 ; $S_+$(BR)= 2.96

\noindent the corresponding predictions for $S_-$ are compatible
---again within the 0.77 renormalization--- with the experimental
 results

$S_-$(exp)= 7.5$\pm$1.2 ; 7.8$\pm$1.9 ;7.5$\pm$0.7 {\it vs.}

$S_-$(KB3)= 7.11 ; $S_-$(BR)= 6.52.

\noindent If we turn to $^{56}$Fe the corresponding numbers are:

$S_+$(exp)= 2.3$\pm$0.6 ; 2.9$\pm$0.3 {\it vs.} $S_+$(KB3)= 2.7, and

$S_-$(exp)= 9.9$\pm$2.4 {\it vs.} $S_-$(KB3)= 9.8.

\noindent We have prefered to omit error bars in the extrapolated
values which are simply reasonable visual guesses. However, they fall
so confortably in the middle of the experimental intervals, that an
estimate of computational uncertainties would leave the conclusions
unchanged.

\medskip

{\bf Comparison of Monte Carlo and Shell Model extrapolations.} The
differences between our results and those of ref.  \cite{dean} are
mostly ---but not only--- due to the use of different forces.  With the
same force the shell model extrapolations yield values that are some
20$\%$ larger than the SMMC ones. The discrepancy is probably related
to the lack of convergence of the SMMC energies detected in table I.

{\em Note } \cite{deanpc}. In the most recent SMMC calculations with
finer $\Delta \beta$ steps of 1/32 (instead of 1/16) the binding
energy goes down by 1 MeV thus eliminating the problem mentioned in
point 2 above.  Furthermore $S_+$ becomes 4.7$\pm$0.3 in full
agreement with our extrapolated value . SMMC values have also become
available for the KB3 interaction \cite{langa}. The values for
$^{54}$Fe and $^{56}$Fe are 6.05$\pm$0.45 and 3.99$\pm$0.27, again in
very good agreement with our values.

\bigskip

\noindent {\bf II. Strength functions: standard quenching and missing
  intensity}

\smallskip In fig.~3 we show the total $l=0$ cross sections obtained
by Anderson {\it et al.} for $^{54}$Fe$(p,n)^{54}$Co. The individual
peaks in table I of \cite{ander} have been associated to gaussians of
$\sigma$=87 keV (the instrumental width) for the lowest and
$\sigma$=141 keV for the others. The ``background'' (the area under
the dashed lines) is obtained by converting the 2 MeV bins in table
III of \cite{ander} into gaussians of $\sigma$=1.41 MeV. Since there
is no direct experimental evidence to decide how much of this
background is genuine strength, two extreme choices are possible to
extract $S_-$ : either keep the whole area in the figure (i.e.,
$S_-=10.3\pm1.4$), or only the area over the dashed line (i.e.,
$S_-=6.0\pm0.4$). An intermediate alternative consists in keeping what
is left of the background after subtracting from it a calculated
contribution to quasi free scattering (QFS).  The resulting profile
(with $S_-=7.5\pm0.7$, the number adopted in I.ii) is shown in fig.~4
(tables I+II of \cite{ander}) and compared with the Lanczos strength
function (see \cite{cpz} for instance) obtained after 60 iterations in
a $t=3$ calculation for the parent state and $t=4$ for the daughters
(the peaks are broadened by gaussians of $\sigma$=87 keV for the
lowest, and $\sigma$=212 keV for the others). The areas under the
measured and calculated curves are taken to be the same (we know that
upon extrapolation to the exact results they coincide). To within an
overall shift of some 2 MeV, the two profiles agree quite nicely.  The
discrepancy is easily traced to the (too large) value of the gap in
table II at this level of truncation.

Although a calculation closer to the exact one would be welcome, the
elements we have point to a situation in all respects similar to that
of the $^{48}$Ca$(p,n)^{48}$Sc reaction, analyzed in \cite{cpz}. What
was shown in this reference can be summed up as follows:

\begin{itemize}
\item The effective $\sigma\tau$ operator to be used in a
  $0\hbar\omega$ calculation is quenched by a factor close to the
  standard one ($\approx 0.77$) through a model independent mechanism
  associated to nuclear correlations.
\item The rest of the strength must be carried by ``intruders'' (i.e.,
  non $0\hbar\omega$ excitations). Only a fraction of this strength is
  located under the resonance, but intruders are conspicuously present
  in this region and make their presence felt through mixing that
  ``dilutes'' the $0\hbar\omega$ peaks causing apparent
  ``background''.
\end{itemize}

In all probability, the long tail in fig.~3 corresponds to intruder
strength and should be counted as such. What is achieved by
subtracting the QFS contribution amounts ---accidentally but
conveniently--- to isolate the $0\hbar\omega$ quenched strength. It is
to this contribution that the notion of standard quenching applies but
it should be kept in mind that the remaining strength ---necessary to
satisfy the $3(N-Z)$ sum rule--- is not missing, but most probably
present in the satellite structure beyond the resonance region as
hinted in the very careful analysis of ref. \cite{ander}.

In fig.~5, to make a meaningful comparison with the
$^{54}$Fe$(n,p)^{54}$Mn data of \cite{romq} the spikes of a $t=3$
calculation have been replaced by gaussians with $\sigma=1.77$ MeV,
chosen to locate some strength at $-$2 MeV, where the first
experimental point is found. The resulting distribution is then
transformed into a histogram with 1 MeV bins.  The agreement is quite
satisfactory. It should be pointed out that the measures of
\cite{vett} are displaced to lower energies by some 700 keV with
respect to those of\cite{romq}.  Otherwise, the experiments are in
good agreement, and both show satellite structure beyond the resonance
(not included in fig.~5, but visible in figs.~10 and 7 in \cite{vett}
and \cite{romq} respectively.

Finally, in figs.~6 and 7 we show the corresponding results for
$^{56}$Fe targets, for which a $t=2$ truncation level was chosen,
going to $t=4$ for $(p,n)$, and $t=2$ for $(n,p)$. Though this
numerical limitation is rather severe, the agreement with the data
remains good enough to support the main conclusion of this paper:

The GT strength functions for $^{54}$Fe and $^{56}$Fe, in the
resonance region and below, are well described by $0\hbar\omega$
calculations that account for $(0.77)^2$ of the total strength. The
remainder, due to intruder states, is likely to be present in the
observed satellite structures, so that the $3(N-Z)$ sum rule is
satisfied.

\smallskip

We have also shown that spurious reductions of the GT+ strength can
occur due to defects of the effective interaction as it is most
probably the case for some of the results of ref. \cite{dean}.

This work has been partly supported by the IN2P3 (France) -- CICYT
(Spain) agreements and by DGICYT(Spain) grant PB93-264.

\begin{table}
\begin{center}
  \leavevmode
\begin{tabular}{|cccc|ccc|}
  $^{54}$Fe & KB3 & BR & E(BR) & $^{56}$Fe & KB3 & BR \\
\hline $t=0$ &
  10.29 & 10.29 & $-50.23$ & & 10.01 & 7.33 \\
  $t=1$ & 9.30 & 9.34 &
  $-51.38$ & & 7.73 & 5.70 \\
  $t=2$ & 7.68 & 7.22 & $-54.67$ & & 6.37 & 4.48\\
  $t=3$ & 7.24 & 6.66 & $-55.30$ & & 5.61 & 3.75 \\
  $t=4$ & 6.70 & 5.84 & $-56.21$ & & 5.11 & \\
  $t=5$ & 6.53 & 5.62 & $-56.48$ & & & \\ SMMC& &
  4.32$\pm$0.24 & $-55.5\pm$0.5 & & & 2.73$\pm$0.04 \\
\end{tabular}
\end{center}
\caption{$^{54}$Fe $\rightarrow  ^{54}$Mn and
$^{56}$Fe $\rightarrow  ^{56}$Mn Gamow Teller strength $S_+$ in
  units of the GT sum rule.
   In column 4 nuclear two body energy of the
  ground state of $^{54}$Fe (in MeV).}
\end{table}

\begin{table}
\begin{center}
\leavevmode
\begin{tabular}{|ccccc|ccccc|}
 KB3 & 3/2 & 5/2 & 1/2 & $\Delta$ &  BR & 3/2 & 5/2 & 1/2 & $\Delta$ \\
\hline
 $t=0$ & 0.0 & 0.38 & 1.15 & 8.57 & & 0.48 & 0.00 & 3.06 & 7.42 \\
 $t=1$ & 0.0 & 0.47 & 1.14 & 7.33 & & 0.07 & 0.00 & 2.11 & 5.80 \\
 $t=2$ & 0.0 & 0.72 & 1.16 & 8.10 & & 0.07 & 0.00 & 2.27 & 7.01 \\
 $t=3$ & 0.0 & 0.76 & 1.14 & 7.74 & & 0.00 & 0.08 & 1.89 & 6.41 \\
 $t=4$ & 0.0 & 0.86 & 1.14 & 7.90 & & 0.00 & 0.11 & 1.83 & 7.21 \\
\hline
 EXP & 0.0 & 0.77 & 1.11 & 6.39 & &      &      &      &      \\
\end{tabular}
\end{center}
\caption{Excitation energies of the low-lying states in
  $^{57}$Ni and the gap $\Delta$ in MeV (see text).  KB3 and BR
  results for several truncations, compared with the experimental
  results.}
\end{table}

\newpage

\begin{figure}
  \begin{center}
    \leavevmode \psfig{file=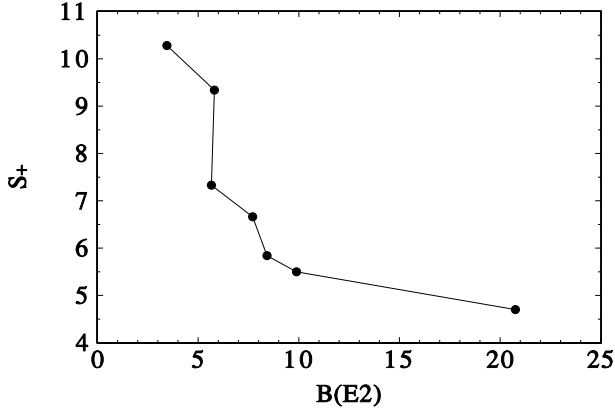,width=8.5cm}
  \end{center}
  \caption{ $^{54}$Fe. $S_+$ (in units of the GT sum rule) {\it vs.} B(E2)
    ($2^{+} \rightarrow 0^{+}$) (in $e^2$$fm^4$).  BR interaction. The
    last point to the right is extracted from the results of
    \protect\cite{alha}.}
\end{figure}

\begin{figure}
  \begin{center}
    \leavevmode \psfig{file=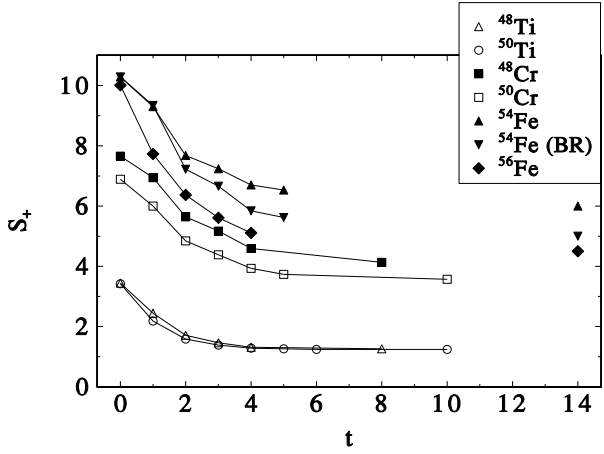,width=8.5cm}
  \end{center}
  \caption{Total GT+ strength as a function of the truncation
    level t. Calculated values are linked by lines, the isolated ones
    are extrapolations.  KB3 interaction, in all cases but one.}
\end{figure}

\begin{figure}
  \begin{center}
    \leavevmode \psfig{file=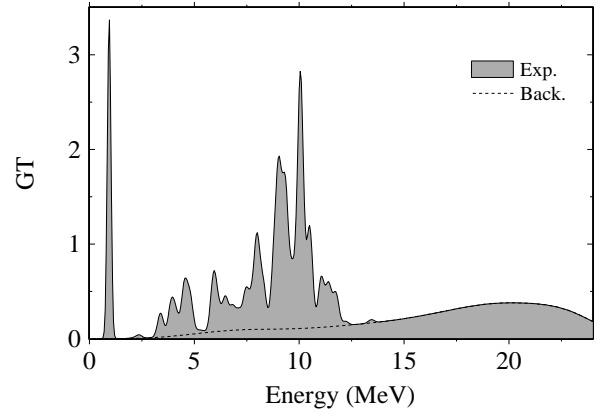,width=8.5cm}
  \end{center}
  \caption{Experimental GT$-$ strength in $^{54}$Fe \protect\cite{ander}}
\end{figure}

\begin{figure}
  \begin{center}
    \leavevmode \psfig{file=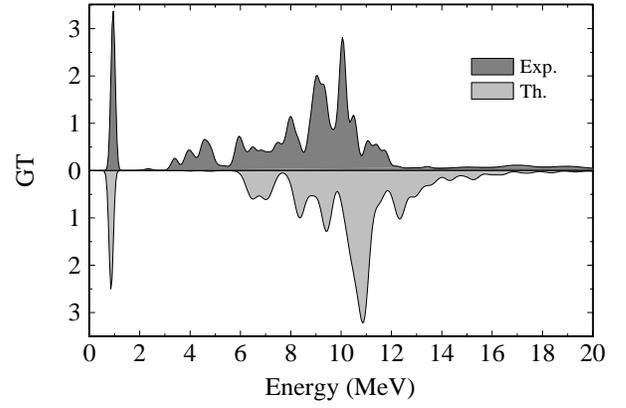,width=8.5cm}
  \end{center}
  \caption{Experimental GT$-$ strength in $^{54}$Fe after QFS subtraction
    \protect\cite{ander}, compared to $t=3 \rightarrow t=4$
    calculations.}
\end{figure}

\newpage

\begin{figure}
  \begin{center}
    \leavevmode \psfig{file=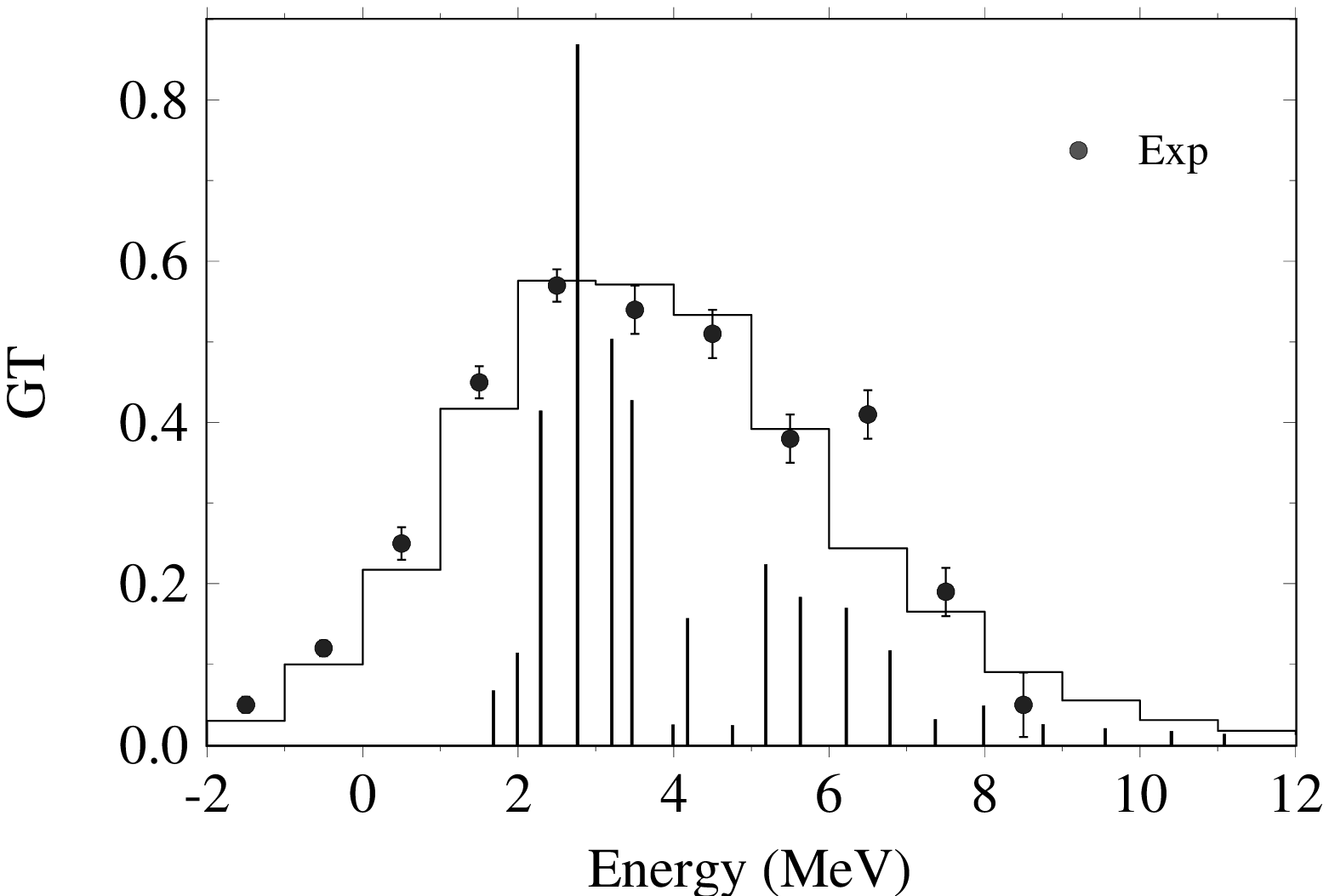,width=8.5cm}
  \end{center}
       \caption{Experimental GT+ strength in $^{54}$Fe
         \protect\cite{romq},compared to $t=3 \rightarrow t=3$
         calculations.}
\end{figure}

\begin{figure}
  \begin{center}
    \leavevmode \psfig{file=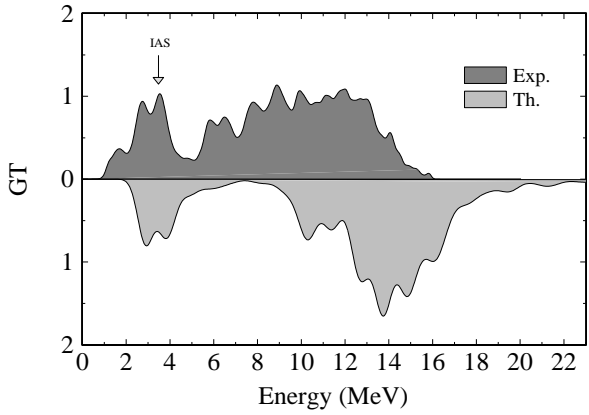,width=8.5cm}
  \end{center}
       \caption{Experimental GT$-$ strength in $^{56}$Fe
         \protect\cite{rapp},compared to $t=2 \rightarrow t=4$
         calculations. The IAS peak has not been removed from the
         data, but it is not included in the calculations.}
\end{figure}

\begin{figure}
  \begin{center}
    \leavevmode \psfig{file=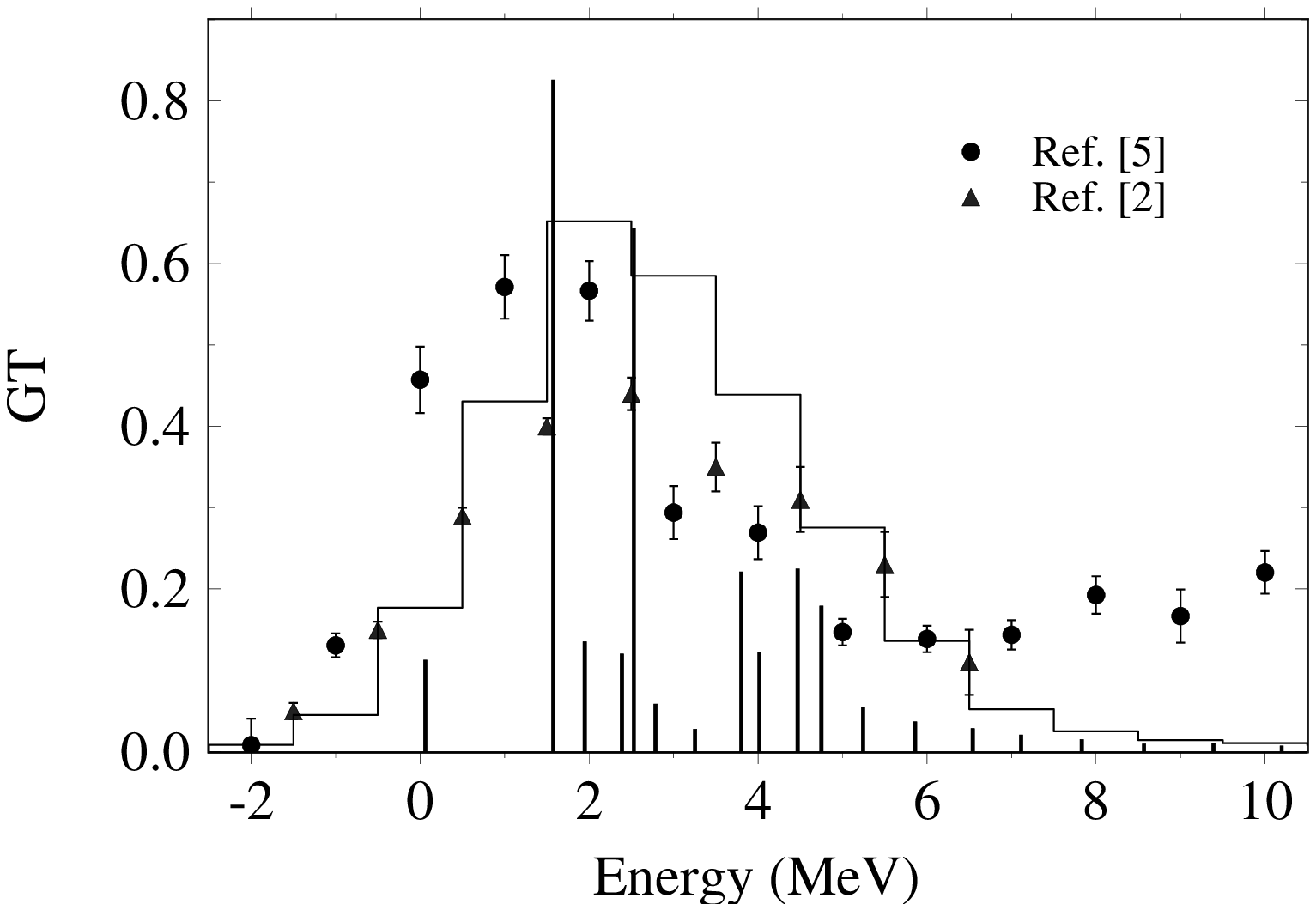,width=8.5cm}
  \end{center}
       \caption{Experimental GT+ strength in $^{56}$Fe
         \protect\cite{romq,elka},compared to $t=2 \rightarrow t=2$
         calculations.}
\end{figure}


\begin{thebibliography}{99}


\bibitem{vett} M.C. Vetterly {\em et al.}, Phys. Rev. C {\bf 40}, 559
  (1989).

\bibitem{romq} T. R\"onnqvist {\em et al.}, Nucl. Phys. {\bf A563}, 225
  (1993).


\bibitem{rapp} J. Rapaport {\em et al.}, Nucl. Phys. {\bf A410}, 371
  (1983).

\bibitem{ander} B. D. Anderson, C. Lebo, A. R. Baldwin, T.
  Chittrakarn, R. Madey, and J. W. Watson, Phys. Rev. C {\bf 41}, 1474
  (1990).

\bibitem{elka} S. El-Kateb {\em et al.}, Phys. Rev. C {\bf 49}, 3120
  (1994).

\bibitem{astro} G. M. Fuller, W. A. Fowler and M. J. Newman, Astophys.
  J.  {\bf 252}, 715 (1982)


\bibitem{bloom} S.D. Bloom and G.M. Fuller, Nucl. Phys. {\bf A440},
  511 (1985).


\bibitem{bint} F. Petrovich, H. Mc Manus, V. A. Madsen and J.
  Atkinson, Phys. Rev. Lett. {\bf 22}, 895 (1969).

\bibitem{auf} M. B. Aufderheide, S. D. Bloom, D. A. Resler, G. J.
  Mathews, Phys. Rev. C {\bf 48}, 1677 (1993).


\bibitem{vanhe} J. F. A. van Hienen, W. Chung and B. H. Wildenthal,
  Nucl. Phys. {\bf A269}, 159 (1976).

\bibitem{koops} J. E. Koops and P. W. M. Glaudemans, Z. Phys. {\bf
    A280}, 181 (1977).

\bibitem{muto} K. Muto, Nucl. Phys. {\bf A451},
  481 (1986).


\bibitem{muint} A. Yokoyama and H. Horie, Phys. Rev. C {\bf 31},1012
  (1985).

\bibitem{auer} N. Auerbach, G. F. Bertsch, B. A. Brown and L. Zhao,
  Nucl. Phys. {\bf A556}, 190 (1993).


\bibitem{brint} W. A. Richter, M. G. van der Merwe, R. E. Julies and
  B. A. Brown, Nucl. Phys. {\bf A523}, 325 (1990).


\bibitem{engel} J. Engel, P. Vogel and M. R. Zirnbauer, Phys. Rev. C
  {\bf 37}, 731 (1988).


\bibitem{alha} Y. Alhassid, D. J. Dean, S. E. Koonin, G. Lang and W.
  E. Ormand, Phys. Rev. Lett. {\bf 72}, 613 (1994).

\bibitem{czpm} E. Caurier, A. P. Zuker, A. Poves and G.
  Mart\'{\i}nez-Pinedo, Phys. Rev. C {\bf 50}, 225 (1994).

\bibitem{kb} T.T.S. Kuo and G.E. Brown, Nucl. Phys. {\bf A114}, 241
  (1968).

\bibitem{pz} A. Poves and A. Zuker, Phys. Rep. {\bf 70}, 235
  (1981).

\bibitem{ant} E. Caurier, code ANTOINE, Strasbourg 1989.


\bibitem{anan} N. Anantaraman {\it et al.}, Phys. Rev. C {\bf 44}, 398
  (1991).

\bibitem{dean} D. J. Dean, P. B. Radha, K. Langanke, Y. Alhassid, S.
  E. Koonin and W. E. Ormand, Phys. Rev. Lett. {\bf 72}, 4066 (1994).

\bibitem{brwil} B. A. Brown and B. H. Wildenthal, At. Data Nucl. Data
  Tables {\bf 33}, 347 (1985).

\bibitem{oster} F. Osterfeld, Rev. Mod. Phys. {\bf 64}, 491 (1992).

\bibitem{cpz} E. Caurier, A. Poves and A. P. Zuker, Phys. Rev. Lett.
  {\bf 74} 1517 (1995).

\bibitem{deanpc} D. J. Dean, private communication.

\bibitem{langa} K. Langanke {\it et al.}, Caltech preprint, march 1995.
  NUC-TH/9504019

\end{thebibliography}
\end{document}